\documentclass[10pt,twoside]{article}
\usepackage{graphicx}
\usepackage{amsmath}
\usepackage{Latex-document}

\title{\bf Approximation Thresholds for \vskip -2mm Combinatorial Optimization
Problems\vskip 6mm}

\author{Uriel Feige\vspace*{-0.5cm}\thanks{Department of Computer Science
and Applied Mathematics, The Weizmann Institute, Rehovot 76100, Israel. E-mail: feige@wisdom.weizmann.ac.il}}
\date{\vspace{-8mm}}

\begin{document}

\maketitle

\thispagestyle{first} \setcounter{page}{649}

\begin{abstract}

\vskip 3mm

An NP-hard combinatorial optimization problem $\Pi$ is said to
have an {\em approximation threshold} if there is some $t$ such
that the optimal value of $\Pi$ can be approximated in polynomial
time within a ratio of $t$, and it is NP-hard to approximate it
within a ratio better than $t$. We survey some of the known
approximation threshold results, and discuss the pattern that
emerges from the known results.

\vskip 4.5mm


\end{abstract}

\vskip 12mm

\section{Introduction}
\label{sec:1}

\vskip-5mm \hspace{5mm}

Given an instance $I$ of a combinatorial optimization problem $\Pi$,
one wishes to find a feasible solution of optimal value to $I$.
For example, in the maximum clique problem, the input instance is
a graph $G(V,E)$, a feasible solution is a clique (a set of vertices
$S \subset V$ such
that for all vertices $u,v \in S$, edge $(u,v)\in E$), the value of the
clique is its size,
and the objective is to find a feasible solution of maximum value.
By introducing an extra parameter $k$ to a combinatorial maximization problem,
one can formulate a decision problem of the form ``does $I$ have a feasible
solution of value at least $k$?'' (or ``at most $k$'', for a minimization
problem). In this work we consider only combinatorial optimization problems
whose decision version is NP-complete. Hence solving these problems is
NP-hard, informally meaning that there is no polynomial time algorithm
that is guaranteed to output the optimal solution for every input instance
(unless P=NP).
(For information on the theory of
NP-completeness, see~\cite{GJ}, or essentially any book on computational
complexity.)

One way of efficiently coping with NP-hard combinatorial optimization problems
is by using polynomial time approximation algorithms. These algorithms
produce a feasible solution that is not necessarily optimal.
The quality of the
approximation algorithm is measured by its approximation ratio: the worst case
ratio between the value of the solution found by the algorithm and the value
of the optimal solution.
Numerous approximation algorithms have been designed for various
combinatorial optimization problems, applying a diverse set of algorithmic
techniques (such as greedy algorithms, dynamic programming, linear programming
relaxations, applications of the probabilistic method, and more).
See for example~\cite{hochbaum,vazirani,compendium}.

When one attempts to design an approximation algorithm for a specific
problem, it is natural to ask whether there
are limits to the best approximation ratio achievable.
The theory of NP-completeness is useful in this context, allowing one to
prove NP-hardness of approximation results. Namely, for some value of $\rho$,
achieving an approximation
ratio better than $\rho$ is NP-hard.
Of particular interest are {\em threshold} results.
A combinatorial optimization problem $\Pi$ is said to have an approximation
threshold at $t$ if
there is a polynomial time algorithm
that approximates the optimal value within a ratio of $t$,
and a hardness of
approximation result that shows that achieving approximation ratios better
than $t$ is NP-hard.

The notion of an approximation threshold is often not fully appreciated,
so let us discuss it in more detail. First, let us make the point that
thresholds of approximation (in the sense above) need not exist at all.
One may well imagine that for certain problems there is a gradual change
in the complexity of achieving various approximation ratios: solving the
problem exactly is NP-hard, approximating it within a ratio of $\rho$
can be done in polynomial time, and achieving approximation ratios between
$\rho$ and~1 is neither in P nor NP-hard, but rather of some intermediate
complexity. The existence of an approximation threshold says that nearly
all approximation ratios (except for ratios that differ from the threshold only
in low order terms)
are either NP-hard to achieve, or in P. This is analogous
to the well known empirical observation that ``most'' combinatorial
optimization
problems that people study turn out to be either in P or NP-hard,
with very few exceptions. But note that in the context of approximation
ratios, the notion of ``most'' is well defined.
The second point that we wish to make is that an NP-hardness of approximation
result is really a polynomial time algorithm. This algorithm reduces
instances of 3SAT (or of some other NP-complete language) to instances
of $\Pi$, and an approximation within a ratio better than $\rho$ for $\Pi$
can be used in order to solve 3SAT. Hence the threshold is a meeting point
between two polynomial time algorithms: the reduction and the approximation
algorithm. Here is another way of looking at it.
For problem $\pi$, we can say that approximation ratio $\rho_1$
reduces to $\rho_2$ if there is a polynomial time algorithm that can
approximate instances of $\pi$ within ratio $\rho_1$ by invoking a subroutine
that approximates instances of $\pi$ within ratio $\rho_2$. (Each call
to the subroutine counts as one time unit.) Two approximation ratios are
equivalent if they are mutually reducible to each other.
The existence of a threshold of approximation for problem $\Pi$ says that
essentially all approximation ratios for it fall into two equivalent classes
(those above the threshold and those below the threshold).
A-priori, it is not clear why the number of equivalent classes should be two.

Hardness of approximation results are often (though not always)
proved using ``PCP techniques''. For more details on these
techniques, see for example the survey of Arora~\cite{arora}. We
remark that in all cases the hardness results apply also to the
problem of {\em estimating} the optimal value for the respective
problem. An estimation algorithm is a polynomial time algorithm
that outputs an upper bound and a lower bound on the value of the
optimal solution (without necessarily producing a solution whose
value falls within this range). The estimation ratio of the
algorithm is the worst case ratio between the upper bound and the
lower bound. An approximation algorithm is stronger than an
estimation algorithm in a sense that it supports its estimate by
exhibiting a feasible solution of the same value. Potentially,
designing estimation algorithms is easier than designing
approximation algorithms. See for example the remark in
Section~\ref{sec:domatic}

For many combinatorial optimization problems, approximation
thresholds are known.
In Section~\ref{sec:survey}
we survey some of these problems.
For each such problem, we sketch an efficient algorithm
for estimating the optimal value
of the solution. The estimation ratios of these algorithms
(the analog of approximation ratios) meet the thresholds of approximation
for the respective problems (up to low order additive terms),
and hence are best possible (unless P=NP).
One would expect these estimation algorithms to be the ``state of the art'' in
algorithmic design.
However, as we shall see, for every problem above there is a
{\em core} version for which the best possible estimation algorithm
is elementary: it bases its estimate only on easily computable properties
of the input instance, such as the number of clauses in a formula,
without searching for a solution.
The core versions that we present often have the property that
Hastad~\cite{hastad2} characterizes
as ``non-approximable beyond the random
assignment threshold''. In these cases (e.g., max 3SAT),
the estimates on the value of
the optimal solution are derived from analysing the expected value
of a random solution.

In contrast, for problems that are known to have a polynomial time
approximation scheme (namely, that can be approximated within ratios
arbitrarily close to 1), their approximation algorithms do perform
an extensive search for a good solution, often using dynamic programming.

The empirical findings are discussed in Section~\ref{sec:discussion}

\section{A survey of some threshold results}
\label{sec:survey}

\vskip-5mm \hspace{5mm}

In this section we survey some of the known threshold of approximation
results. For each problem we show simple lower bounds and
upper bounds on
the value of the optimal solution. Obtaining any better bounds is NP-hard.
(There are certain exceptions to this. For set cover and domatic number
the matching hardness result are under the assumption that NP does not
have slightly super-polynomial time algorithms. The hardness results for
clique and chromatic number
assume that NP does not have randomized algorithms that run
in expected polynomial time.).

{\bf Conventions.} For a graph, $n$ denotes the number of vertices, $m$ denotes
the number of edges, $\delta$ denotes the minimum degree, $\Delta$ denotes
the maximum degree. For a formula, $n$ denotes the number of variables,
$m$ denotes the number of clauses. For each problem we define
its inputs, feasible solutions, value of solutions, and objective.
We present the known approximation ratios and hardness results, which
are given up to low order additive terms.
We then present a {\em core} version of the problem. For the core version,
we present an upper bound and a
lower bound on the value of the optimal solution.
Improving over these bounds (in the worst case) is NP-hard.
In most cases, we give hints to the proofs of our claims. We also
cite references were full proofs can be found.

\subsection{Max coverage~\cite{johnson,setcover}}

\noindent
{\bf Input.} A collection $S_1, \ldots, S_m$ of sets with
$\bigcup_{i=1}^m S_i = \{1, \ldots, n\}$. A parameter $k$.

\noindent {\bf Feasible solution.} A collection $I$ of $k$
indices.

\noindent {\bf Value.} Number of covered items, namely
$|\bigcup_{i\in I} S_i|$.

\noindent
{\bf Objective.} Maximize.

\noindent {\em Algorithm.} Greedy. Iteratively add to $I$ the set
containing the maximum number of yet uncovered items, breaking
ties arbitrarily.

\noindent
{\em Approximation ratio.} $1 - 1/e$.

\noindent
{\em Hardness.} $1 - 1/e$. 

\noindent
{\bf Core.}
$d$-regular, $r$-uniform.
Every set is of cardinality $d$. Every item is in
$r$ sets. $k = n/d$.

\noindent
{\em Upper bound.} $n$.

\noindent
{\em Lower bound.} $(1 - 1/e)n$. (Pick $k = {n/d} = {m/r}$ sets at random.)

\noindent
{\em Hardness of core.} $(1 - 1/e + \epsilon)$ for every
$\epsilon > 0$, when $d,r$ are large enough.

\noindent {\bf Remarks.} For the special case in which each item
belongs to exactly~4 sets, and $k = m/2$, there always is a choice
of sets covering at least $15n/16$ items. For very $\epsilon > 0$,
a $(1 + \epsilon)15/16$ approximation ratio is
NP-hard~\cite{holmerin}.

\subsection{Min set cover~\cite{johnson,setcover}}

\noindent
{\bf Input.} A collection $S_1, \ldots, S_m$ of sets with
$\bigcup_{i=1}^m S_i = \{1, \ldots, n\}$.

\noindent
{\bf Feasible solution.} A set of indices $I$ such that
$\bigcup_{i\in I} S_i = \{1, \ldots, n\}$.

\noindent
{\bf Value.} Number of sets used in the cover, namely, $|I|$.

\noindent
{\bf Objective.} Minimize.

\noindent {\em Algorithm.} Greedy. Iteratively add to $I$ the set
containing the maximum number of yet uncovered items, breaking
ties arbitrarily.

\noindent
{\em Approximation ratio.} $\ln n$. 

\noindent
{\em Hardness.} $\ln n$. 

\noindent
{\bf Core.}
$d$-regular, $r$-uniform.
Every set is of cardinality $d$. Every item is in
$r$ sets.

\noindent
{\em Lower bound.} $n/d$.

\noindent
{\em Upper bound.} $(n/d) \ln n$, up to low order terms.
(Include every set in $I$ independently
with probability $\frac{\ln n}{r}$.)

\noindent
{\em Hardness of core.} $\ln n$. 

\noindent {\bf Remarks.} The hardness results for set cover
in~\cite{setcover} assume that NP does not have deterministic
algorithms that run in slightly super-polynomial time (namely,
time $n^{O(\log \log n)}$).

\subsection{Domatic number~\cite{FHKS}}
\label{sec:domatic}

\noindent
{\bf Input.}
A graph.

\noindent
{\bf Feasible solution.} A domatic partition of the graph. That is, a partition
of the vertices of the graph into disjoint sets, where each set is
a dominating set. (A dominating set is a set $S$ of vertices that is adjacent
to every vertex not in $S$.)

\noindent
{\bf Value.} Number of dominating sets in the partition.

\noindent
{\bf Objective.} Maximize.

\noindent {\em Algorithm.} Let $\delta$ be the minimum degree in
the graph. Partition the vertices into $(1 - \epsilon)(\delta +
1)/\ln n$ sets at random, where $\epsilon$ is arbitrarily small
when $\delta/\ln n$ is large enough. Almost all these sets will be
dominating. The sets that are not dominating can be unified with
the first of the dominating sets to give a domatic partition. The
algorithm can be derandomized to give (a somewhat unnatural)
greedy algorithm.

\noindent
{\em Approximation ratio.} $1/\ln n$.

\noindent
{\em Hardness.} $1/\ln n$. 

\noindent
{\bf Core.} $\delta/\ln n$ is large enough.

\noindent
{\em Upper bound.} $\delta + 1$.

\noindent
{\em Lower bound.} $(1 - \epsilon)(\delta + 1)/\ln n$.

\noindent
{\em Hardness of core.}
$(1 + \epsilon)/\ln n$, for every $\epsilon > 0$.

\noindent {\bf Remarks.} The hardness results assume that NP does
not have deterministic algorithms that run in time $n^{O(\log \log
n)}$. The lower bound can be refined to $(1 - \epsilon)(\delta +
1)/\ln \Delta$, where $\Delta$ is the maximum degree in the graph.
This is shown using a nonconstructive argument (based on a two
phase application of the local lemma of Lovasz). It is not known
how to find such a domatic partition in polynomial time.

\subsection{$k$-center~\cite{HN,DF,HS}}

\noindent
{\bf Input.} A metric on a set of $n$ points
and a parameter $k < n$.

\noindent
{\bf Feasible solution.} A set $S$ of $k$ of the points.

\noindent
{\bf Value.} Distance between $S$ and point furthest away from $S$.

\noindent
{\bf Objective.} Minimize.

\noindent
{\em Algorithm.} Greedy.
Starting with the empty set, iteratively add into $S$
the point furthest away from $S$, resolving ties arbitrarily.

\noindent
{\em Approximation ratio.} 2.

\noindent
{\em Hardness.} 2.

\noindent
{\bf Core.}
The $n$ points are vertices of a
graph with minimum degree $\delta$,
equipped with a shortest distance metric.
$k \ge n/(\delta+1)$.

\noindent
{\em Lower bound.} 1. (Because $k < n$.)

\noindent
{\em Upper bound.} 2. (As long as there is a vertex of distance more than~2
from the current set $S$,
every step of the greedy algorithm covers at least $\delta+1$ vertices.)

\noindent
{\em Hardness of core.} 2.
(Because it is
NP-hard to tell if the underlying graph has a dominating set of size $k$.)

\subsection{Min sum set cover~\cite{FLT}}

\noindent
{\bf Input.} A collection $S_1, \ldots, S_m$ of sets with
$\bigcup_{i=1}^m S_i = \{1, \ldots, n\}$.

\noindent
{\bf Feasible solution.}
A linear ordering. That is, a one to one
mapping $f$ from the collection of sets to $\{1, \ldots, m\}$.

\noindent
{\bf Value.} The average time by which an item is covered.
Namely, $\frac{1}{n}\sum_i \min_{\{j|i \in S_j\}} f(j)$.

\noindent
{\bf Objective.} Minimize.

\noindent
{\em Algorithm.} Greedy. Iteratively pick the set containing the
maximum number of yet uncovered items, breaking ties arbitrarily.

\noindent
{\em Approximation ratio.} 4.

\noindent
{\em Hardness.} 4.

\noindent
{\bf Core.}
$d$-regular, $r$-uniform.
Every set is of cardinality $d$. Every item is in
$r$ sets.

\noindent
{\em Lower bound.} $n/2d = m/2r$. (Because every set covers $d$ items.)

\noindent
{\em Upper bound.} $m/(r+1)$. (By ordering the sets at random.)

\noindent
{\em Hardness of core.} $2 - \epsilon$, for every $\epsilon > 0$,
when $r$ is large enough.

\noindent {\bf Remarks.} Note that the core has a lower threshold
of approximation than the general case.

\subsection{Min bandwidth~\cite{unger}}

\noindent
{\bf Input.}
A graph.

\noindent
{\bf Feasible solution.}
A linear arrangement. That is, a one to one
mapping $f$ from the set of
vertices to $\{1, \ldots, n\}$.

\noindent
{\bf Value.} Longest stretch of an edge.
Namely, the maximum over all edges $(i,j)$ of $|f(i) - f(j)|$.

\noindent
{\bf Objective.} Minimize.

\noindent
{\bf Core.}
The input graph $G$ is a unit length circular arc graph.
Namely, the vertices represent arcs of equal length on a circle.
Two vertices are connected by an edge if their respective arcs overlap.
Let $\omega(G)$ denote the size of the maximum clique
in $G$. (In circular arc graphs,
a clique corresponds to
a set of mutually
intersecting arcs, and $\omega(G)$ can be computed easily.)

\noindent
{\em Lower bound.}
$\omega(G) - 1$.

\noindent
{\em Upper bound.} $2\omega(G) - 2$.

\noindent
{\em Hardness of core.} 2. 

\noindent {\bf Remarks.} No threshold of approximation is known
for the problem on general graphs.

\subsection{Max 3XOR~\cite{hastad2}}

\noindent
{\bf Input.}
A logical formula with $n$ variables and $m$ clauses.
Every clause is the {\em exclusive or} of three distinct literals.

\noindent
{\bf Feasible solution.} An assignment to the $n$ variables.

\noindent
{\bf Value.} Number of clauses that are satisfied.

\noindent
{\bf Objective.} Maximize.

\noindent
{\em Algorithm.} Gaussian elimination can be used in order to test if the
formula is satisfiable. If the formula is not satisfiable, pick a random
assignment to the variables.

\noindent
{\em Approximation ratio.} $1/2$. (In expectation a random assignments
satisfies $m/2$ clauses, whereas no assignment satisfies more
than $m$ clauses.)

\noindent
{\em Hardness.} $1/2$.

\noindent
{\bf Core.}
Same as above.

\noindent
{\em Upper bound.}
$m$.

\noindent
{\em Lower bound.}
$m/2$.

\noindent
{\em Hardness of core.}
$1/2$. 


\subsection{Max 3AND~\cite{hastad2,zwick}}

\noindent
{\bf Input.}
A logical formula with $n$ variables and $m$ clauses.
Every clause is the {\em and} of three literals.

\noindent
{\bf Feasible solution.} An assignment to the $n$ variables.

\noindent
{\bf Value.} Number of clauses that are satisfied.

\noindent
{\bf Objective.} Maximize.

\noindent
{\em Algorithm.}
Based on semidefinite programming.

\noindent
{\em Approximation ratio.} $1/2$.

\noindent
{\em Hardness.} $1/2$. 

\noindent
{\bf Core.} Pairwise independent version.
Every variable appears at most once in each clause.
For a pair of variables
$x_i$ and $x_j$, let $n_{ij}(0,0)$
($n_{ij}(0,1)$, $n_{ij}(1,0)$, $n_{ij}(1,1)$, respectively) denote the number
of clauses in which they both appear negated ($i$ negated and $j$ positive,
$i$ positive and $j$ negated, both positive, respectively).
Then for every $i$ and $j$,
$n_{ij}(0,0) = n_{ij}(0,1) = n_{ij}(1,0) = n_{ij}(1,1)$.

\noindent
{\em Upper bound.} $m/4$. (Consider the pairs of literals in the satisfied
clauses. There must be at least
three times as many pairs in unsatisfied clauses.)

\noindent
{\em Lower bound.} $m/8$. (The expected number of clauses satisfied by a
random assignment.)

\noindent
{\em Hardness of core.} $1/2$.

\noindent {\bf Remarks.} This somewhat complicated core comes up
as an afterthought, by analysing the structure of instances that
result from the reduction from max 3XOR.

\subsection{Max 3SAT~\cite{hastad2,KZ}}

\noindent
{\bf Input.}
A logical formula with $n$ variables and $m$ clauses.
Every clause is the {\em or} of three literals.

\noindent
{\bf Feasible solution.} An assignment to the $n$ variables.

\noindent
{\bf Value.} Number of clauses that are satisfied.

\noindent
{\bf Objective.} Maximize.

\noindent
{\em Algorithm.}
Based on semidefinite programming.

\noindent
{\em Approximation ratio.} $7/8$.

\noindent
{\em Hardness.} $7/8$. 

\noindent
{\bf Core.} In every clause, the three literals are distinct.

\noindent
{\em Upper bound.} $m$.

\noindent
{\em Lower bound.} $7m/8$. (The expectation for a random assignment.)

\noindent
{\em Hardness of core.} $7/8$. 

\noindent {\bf Remarks.} The approximation ratio of the algorithm
of~\cite{KZ} is determined using computer assisted analysis.

\subsection{Inapproximable problems}
\vskip -5mm \hspace{5mm}

For some problems the best approximation ratios known are of the form
$n^{1 - \epsilon}$ for every $\epsilon > 0$ (where the range of the
objective function is between~1 and $n$). This is often interpreted as
saying that approximation algorithms are almost helpless with respect
to these problems. Among these problems we mention
finding the smallest maximal independent set~\cite{hall},
max clique~\cite{hastad1}, and chromatic number~\cite{FK}.
The hardness results in~\cite{hastad1,FK} are proved under the assumption
that NP does not have expected polynomial time randomized algorithms.

\subsection{Thresholds within multiplicative factors}
\vskip -5mm \hspace{5mm}

For some problems, the best approximation ratio is of the form $O(n^c)$,
for some $0 < c < 1$, and there is a hardness of approximation result
of the form $\Omega(n^{c'})$, where $c'$ can be chosen arbitrarily close
to $c$. We may view these as thresholds of approximation up to a low
order multiplicative factor. An interesting example of this sort
is the {\em max disjoint paths} problem~\cite{GKRSY}.

\noindent
{\bf Input.} A directed graph and a set $S$ of pairs of terminals
$\{(s_1,t_1), \ldots (s_k,t_k)\}$.

\noindent
{\bf Feasible solution.} A collection of edge disjoint paths, each connecting
$s_i$ to $t_i$ for some $i$.

\noindent
{\bf Value.} Number of
pairs of terminals from $S$ that are connected by some path in the solution.

\noindent
{\bf Objective.} Maximize.

\noindent
{\em Algorithm.} Greedy. Iteratively add the shortest path that connects
some yet unconnected pair from $S$.

\noindent
{\em Approximation ratio.} $m^{-1/2}$, where $m$ is the number of arcs
in the graph.

\noindent
{\em Hardness.} $m^{-1/2 + \epsilon}$, for every $\epsilon > 0$.

\noindent
{\bf Core.}
There is a path from $s_1$ to $t_1$.

\noindent
{\em Upper bound.} $k$. At most all pairs can be connected
simultaneously by disjoint paths.

\noindent
{\em Lower bound.} 1. There is a path from $s_1$ to $t_1$.

\noindent
{\em Hardness of core.} $1/k$.
Here $k$ can be chosen as $m^{1/2 - \epsilon}$ for arbitrarily small
$\epsilon$.

\section{Discussion}
\label{sec:discussion}

\vskip-5mm \hspace{5mm}

We summarized the pattern presented in Section~\ref{sec:survey} Given an NP-hard combinatorial optimization
problem that has an approximation threshold, we identify a core version of the problem. For the core version we
identify certain key parameters. Thereafter, an upper bound and a lower bound on the value of the optimal solution
is expressed by a formula involving only these parameters. Even an algorithm that examines the input for
polynomial time cannot output better lower bounds or upper bounds on the value of the optimal solution (in the
worst case, and up to low order terms), unless P=NP.

Note that if we do not restrict what qualifies as being
a key parameter, then the pattern above can be enforced on essentially
any problem with an approximation threshold.
We can simply take the key parameter to be the output of an approximation
algorithm for the same problem. Likewise, if we do not restrict what
qualifies as a core version, we can simply take the core version
to be all those input instances on which a certain approximation algorithm
outputs a certain value.

Hence we would like some restrictions on what may qualify as a key parameter,
or as a core version. One option is to enforce a computational complexity
restriction. Namely, the core should be such that
deciding whether an input instance belongs to the core
is computationally easy, for example, computable in logarithmic space.
(Note that in this respect, the core that we defined for the max disjoint
paths problem may be problematic, because testing whether there is a directed
path from $s_1$ to $t_1$ is complete for nondeterministic logarithmic space.)
Likewise, computing the key parameters should be easy.
Another option is to enforce structural restrictions.
For example, membership in the core should be invariant over renaming
of variables. Ideally, the notions of ``core'' and ``key parameter''
should be defined well enough so that we should be able to say that certain
classes of inputs do not qualify as being a core (e.g.,
because the class is not closed under certain operations),
and that certain parameters
are not legitimate parameters to be used by an estimation algorithm.
Also, the definitions should allow in principle the possibility that
certain problems have approximation thresholds without having a core.

Hand in hand with suggesting formal definitions to the notion of a core,
it would be useful to collect more data. Namely, to find more approximation
threshold results, and to identify plausible core versions to these problems.
As the case of min bandwidth shows, one may find core versions even for
problems for which an approximation threshold is not known.

In this manuscript we mainly addressed the issue of collecting data regarding
approximation thresholds, and describing this data using the notion
of a core. The issue of uncovering principles that explain why the patterns
discussed in this manuscript emerge is left to the reader.

\label{lastpage}

\end{document}